\documentclass[10pt]{article}

\usepackage{latexsym}

\usepackage{graphicx}
\usepackage{bm}
\usepackage{epstopdf}
\usepackage{amsmath,amsfonts,paralist}

\usepackage{amsthm, mathtools}

\newtheorem{theorem}{Theorem}[section]

\usepackage{subfig}

\title{Replica Theory and Spin Glasses}

\author{ F. Morone$^{1}$, F.  Caltagirone$^{2}$, Elizabeth Harrison$^{3}$ and G. Parisi$^{4}$\\ 
\\
\normalsize $^1$ Levich Institute and Physics Department, City College of New York, New York, NY 10031, USA \\ 
\normalsize $^2$ Institut de Physique Th\'eorique, CEA Saclay, 91191 Gif-sur-Yvette, France \\
\normalsize $^3$ Aston University, Aston Triangle, Birmingham B4 7ET, UK \\
\normalsize $^4$ Dip. di Fisica Universit\`a ``La Sapienza'' and INFN, Piazzale A. Moro 2, I-00185, Rome, Italy}

\date{}

\begin{document}

\maketitle

\begin{abstract}
These are notes from the lectures of Giorgio Parisi given at the autumn school ``Statistical Physics, Optimization, Inference, and Message-Passing Algorithm'', that took place in Les Houches, France from Monday September 30th, 2013, till Friday October 11th, 2013. The school was organized by Florent Krzakala from UPMC and ENS Paris, Federico Ricci-Tersenghi from ``La Sapienza'' Roma, Lenka Zdeborov\'a from CEA Saclay and CNRS, and Riccardo Zecchina from Politecnico Torino. \\
\\
The first lecture contains an introduction to the replica method, along with a concrete application to the computation of the eigenvalue distribution of random matrices in the GOE.
In the second lecture, the solution of the SK model is derived, along with the phenomenon of replica symmetry breaking (RSB). 
In the third part, the physical meaning of the RSB is explained. The ultrametricity of the space of pure states emerges as a consequence of the hierarchical RSB scheme. Moreover, it is shown how some low temperature properties of physical observables can be derived by invoking the stochastic stability principle. 
Lecture four contains some rigorous results on the SK model: the existence of the thermodynamic limit, and the proof of the exactness of the hierarchical RSB solution.  
\end{abstract}

\tableofcontents

\section{Introduction to the replica method: the Wigner law}

We introduce the basic idea of the replica method by computing the spectrum of random matrices. To be concrete, we explore the case of real symmetric matrices of large order $N$ having random elements $J_{ij}$ that are independently distributed with zero mean and variance $\overline{J^2}=N^{-1}$. We will always denote the average over the $J$'s with an overbar.  This is problem is widely studied: many rigorous results are known and many different techniques	can be used to compute the spectrum of these matrices.

We also assume that  $\ \overline{J^4}\big/\big(\overline{J^2}\big)^2<\infty\ $ (i.e. the probability distribution of $J$ has no fat tails): less stringent conditions may be imposed, but we are not interested here to find the optimal condition.

Before describing the replica method, we will perform the calculations using a probabilistic approach, which is at the heart of the cavity method.
\subsection{Computing the spectrum with cavity method}

The resolvent $\hat{R}$ of the $N\times N$ matrix $\hat{J}$ is defined as follows 
\footnote{Here we use the convention of denoting the matrices with an hat: ' $\hat{}$ ', while, when we refer to particular matrix elements, we omit the 'hat' symbol.} 
\begin{equation}
\hat{R}(\mathcal{E})=\left(\mathcal{E}\mathbb{I}-\hat{J}\right)^{-1}\ ,
\end{equation}
where $\mathbb{I}$ is the $N\times N$ identity matrix and $\mathcal{E}$ is a complex number (only at the end of the computation we shall take the limit where $\mathcal{E}$ is real.
 
We are interested in computing the trace of $\hat{R}(\mathcal{E})$:
\begin{equation}
\mathrm{Tr}\big[\hat{R}(\mathcal{E})\big]\ =\ \sum_{k}\frac{1}{\mathcal{E}-\mathcal{E}_k}\ ,
\end{equation}
 in the large $N$ limit. Many physical interesting quantities may be extracted from the knowledge of the trace of the resolvent, the simplest one (and the only one we shall consider here)  is the spectral density.
 
As far as $\hat{J}$ is a random matrix, also the trace of the resolvent is a random quantity. We will proceed by first writing exact relations at fixed $\hat{J}$  and at a later stage we shall perform the average for the random matrices.

The idea behind the cavity computation consists in finding a recursion equation between $\mathrm{Tr}\big[\hat{R}^{(N)}\big]$ and $\mathrm{Tr}\big[\hat{R}^{(N+1)}\big]$, where $\hat{R}^{(N)}$ is the $N\times N$ resolvent matrix.  In order to do this it is sufficient to write the matrix element $R^{(N+1)}_{N+1,N+1}$ as a function of the matrix elements $R^{(N)}_{ij}$. 

Let us define for simplicity the following $N \times N$ matrix $\hat{M}^{(N)}\equiv\mathcal{E}\mathbb{I}-\hat{J}$, where $J$ is a matrix with dimensions larger that $N+1$ and the equality holds only in the $N$-dimensional space. It is easy to verify that
\begin{equation}
 R^{(N+1)}_{N+1,N+1}\ =\ \frac{\det[\hat{M}^{(N)}]}{\det[\hat{M}^{(N+1)}]}\ .
\end{equation}
 On the other hand the determinant $\det[\hat{M}^{(N+1)}]$ can be Laplace expanded first along the last row, and then on the last column, thus obtaining:
\begin{equation}
\det[\hat{M}^{(N+1)}]\ =\ M^{(N+1)}_{N+1,N+1}\det[\hat{M}^{(N)}]\ -\ \sum_{k,\ell=1}^{N}M^{(N+1)}_{N+1,k}\ 
M^{(N+1)}_{\ell,N+1}\ C^{(N)}(\ell,k)\ ,
\end{equation}
where $C^{(N)}(\ell,k)$ is the $(\ell,k)$ cofactor of the matrix $\hat{M}^{(N)}$. Dividing the previous expression by $\det[\hat{M}^{(N)}]$ and recalling the definition of $\hat{M}^{(N)}$, we get
\begin{equation}
\frac{1}{R^{(N+1)}_{N+1,N+1}}\ =\ \mathcal{E}\ -\ J^{(N+1)}_{N+1,N+1}\ -\ \sum_{k,\ell=1}^{N}J^{(N+1)}_{N+1,k}\ 
J^{(N+1)}_{\ell,N+1}\ R^{(N)}_{k\ell}\ .\label{eq:recursion1}
\end{equation}

Now we must do the crucial assumption that the off-diagonal elements of the resolvent are of order $ O\left(N^{-1/2}\right)$ (as we shall see later this assumption is the moral equivalent of the validity of replica symmetry). The motivation for this assumption (that can be rigorously proved) are the following:
\begin{itemize}
\item For large values of $\mathcal{E}$ the resolvent can be expanded in inverse power of $\mathcal{E}$ and it possible to check at each order in this expansion that he off diagonal elements of the resolvent are of order $ O\left(N^{-1/2}\right)$.
\item If we assume that $\overline{\left|R^{(N)}_{k\ell}\right|^2}=A/N$, a computation similar to the previous one (going from $N$ to $N+2$) predicts consistently  a finite value for $A$, at least for $\mathcal{E}$ not on the real axis. (In the replica language this statement is equivalent to the condition of stability of the replica symmetric fixed point).
\end{itemize}

If we use the assumption that the off-diagonal elements of the resolvent are of order $ O\left(N^{-1/2}\right)$, in the large $N$ limit the recursion equation \eqref{eq:recursion1} becomes
\begin{equation}
R^{(N+1)}_{N+1,N+1}=\frac{1}{\mathcal{E}\ -\ N^{-1}\sum_{k=1}^{N} 
 R^{(N)}_{kk}}\ +\ O\left(N^{-1/2}\right)\ .\label{eq:recursion2}
\end{equation}
In the last step we have used the fact that $J^{(N+1)}$ and $R^{(N)}$ are uncorrelated quantities.

At this point, a little bit of thought should convince oneself that the matrix elements $R^{(N+1)}_{N+1,N+1}$ and $R^{(N)}_{kk}$ are all identically distributed as a consequence of the fact that the $J$'s are identically distributed and uncorrelated. Taking the average on both sides of Eq. \eqref{eq:recursion2}, together with the limit $N\to\infty$, 
we obtain the following fixed point equation 
\begin{equation}
R(\mathcal{E})\ =\ \frac{1}{\mathcal{E}-R(\mathcal{E})}\ ,\label{eq:FPeq}
\end{equation}
where $R(\mathcal{E})\ \equiv\ \overline{R_{ij}(\mathcal{E})}$. Moreover that dependence on $k$ of $R^{(N)}_{kk}$ should disappear in the limit where $N$ goes to infinity and the diagonal element do not fluctuate. 

The solution   to Eq. \eqref{eq:FPeq} is given by
\begin{equation}
R(\mathcal{E})=\frac{\mathcal{E}\ \pm\ \sqrt{\mathcal{E}^2-4}}{2}\ .\label{eq:AsymptResol}
\end{equation}
Which of the two determinations of the square root should we take? Let us consider what happens on the real line: we must have that  for large $|\mathcal{E}|$, $R(\mathcal{E})$ goes to zero as $1/R(\mathcal{E})$. Therefore in the region $\mathcal{E}>2$ we must take the negative sign, while for $\mathcal{E}<-2$ we must take the positive sign. This may look strange; however the function  $R(\mathcal{E})$ has a cut on the real axis from $-2$ to 2: the choice of the determination for negative values of  $\mathcal{E}$  is exactly what we get if we start from $\left(\mathcal{E}\ \pm\ \sqrt{\mathcal{E}^2-4}\right)/{2}$ and we perform an analytic continuation from positive to negative $\mathcal{E}$, avoiding the cut on the real axis.

Let us now introduce the density of states $\rho(\lambda)$, which is defined as follows
\begin{equation}
\rho(\lambda)\ \equiv\ \frac{1}{N}\sum_{k=1}^{N}\ \delta(\lambda-\mathcal{E}_k)\ .
\end{equation}
By using the density of states $\rho(\lambda)$, we can rewrite the trace of $\hat{R}(\mathcal{E})$ as
\begin{eqnarray}
\nonumber
\mathrm{Tr}\big[\hat{R}(\mathcal{E})\big]\ &&=\ \sum_{k=1}^{N}\frac{1}{\mathcal{E}-\mathcal{E}_k}\ =\ 
N\lim_{\epsilon\to0}\int\mathrm{d}\lambda\ \frac{\rho(\lambda)}{\mathcal{E}-\lambda-i\epsilon}\ = \\
&&\ N\left[\int\mathrm{d}\lambda\ \rho(\lambda)\ \mathrm{P}\left(\frac{1}{\mathcal{E}-\lambda}\right)+i\pi\rho(\mathcal{E})\right]\ ,
\end{eqnarray}
where the distribution $\mathrm{P}(1/x)$ denotes the principal value of $1/x$. Now,
since $\overline{\mathrm{Tr}\big[\hat{R}(\mathcal{E})\big]}=NR(\mathcal{E})$, we can relate the density of states $\rho(\lambda)$ to the imaginary part of $R(\mathcal{E})$ via the following equation:
\begin{equation}
\rho(\mathcal{E})\ =\ \frac{1}{\pi}\ \mathrm{Im}\left[R(\mathcal{E})\right]\ ,
\end{equation}
from which we get the Wigner semicircle law:
\begin{equation}
\rho(\mathcal{E})=\left(\frac{1}{2\pi}\ \sqrt{4-\mathcal{E}^2}\right)\Theta(2-|\mathcal{E}|)\ .\label{eq:Wigner}
\end{equation}
In the nutshell the function $R(\mathcal{E})$ is a real analytic function with a cut on the real line and the discontinuity on the cut is the density of states. It should be stressed that these results are valid for the resolvent when we are a little away from the cut. Near the cut new effects arise and the computation become more involved.

\subsection{Computing the spectrum with replicas}
We now perform the same calculation using the replica method. The object we are interested in is the average of the trace of the resolvent for an $N\times N$ random matrix, which can be written as:
\begin{equation}
\overline{\mathrm{Tr}\big[\hat{R}(\mathcal{E})\big]}\ =\ \overline{\mathrm{Tr}\left(\frac{1}{\mathcal{E}\mathbb{I}-\hat{J}}\right)}\ =\ 
\frac{\mathrm{d}}{\mathrm{d}\mathcal{E}}\ \overline{\mathrm{Tr}\ \log\left(\mathcal{E}\mathbb{I}-\hat{J}\right)}\ =\ 
\frac{\mathrm{d}}{\mathrm{d}\mathcal{E}}\ \overline{\log\det\left(\mathcal{E}\mathbb{I}-\hat{J}\right)}\ .\label{eq:TraceResolvent}
\end{equation}
Now we can write the average of the logarithm of the determinant in the following way:
\begin{equation}
 \overline{\log\det\left(\mathcal{E}\mathbb{I}-\hat{J}\right)}\ = -2\lim_{n\to 0}\ \frac{\mathrm{d}}{\mathrm{d}n}\ \overline{\left(\frac{1}{\det(\mathcal{E}\mathbb{I}-\hat{J})}\right)^{n/2}}\ .\label{eq:Introreplica}
\end{equation}
This is the famous replica trick. Operatively one computes the quantity on the r.h.s for integer (positive) values of the \textit{replica} number $n$, then one performs an analytical continuation of the result to real values of $n$ and eventually takes the limit $n\to 0$. The main assumption concerning the analytical continuation relies on the possibility of uniquely determining an analytic function knowing it only on a countable (infinite) set of points (the positive integers in our case). This is not a real point of concern: the problems will arise in exchanging the limits $N\to\infty$ and $n\to 0$.

To go further in the calculation, we represent the determinant in the r.h.s of Eq. \ref{eq:Introreplica} using a Gaussian integral:
\begin{equation}
\begin{aligned}
&\overline{\left(\frac{1}{\det(\mathcal{E}\mathbb{I}-\hat{J})}\right)^{n/2}}\ =\ 
\overline{
\int\left(\prod_{i=1}^{N}\prod_{a=1}^{n}\ \frac{\mathrm{d}\phi_i^a}{\sqrt{2\pi}}\right)\exp\left[-\frac{1}{2}\sum_{i,k=1}^{N}\sum_{a=1}^{n}\phi_i^a\left(\mathcal{E}\delta_{ik}-J_{ik}\right)\phi_k^a\right]}=\\
=&\ 
\int\left(\prod_{i=1}^{N}\prod_{a=1}^{n}\ \frac{\mathrm{d}\phi_i^a}{\sqrt{2\pi}}\right)\exp\left[-\frac{1}{2}\sum_{i=1}^{N}\sum_{a=1}^{n}\mathcal{E}\ (\phi_i^a)^2-
\frac{1}{4N}\sum_{i,k=1}^{N}\sum_{a,b=1}^{n}\phi_i^a\phi_k^a\phi_i^b\phi_k^b\right]\ , \label{eq:AverageOverReplicas}
\end{aligned}
\end{equation}
where terms of order $O(1)$ in the argument of the exponential have been neglected. Notice that, if the distribution of the matrix elements were a Gaussian one, the previous representation would be exact.

In order to decouple the sites $i$, we introduce the following \textit{order parameter}:
\begin{equation}
q_{ab}=\frac{1}{N}\sum_{i=1}^{N}\phi_i^a\phi_i^b\ .\label{eq:overlap}
\end{equation}
Inserting a delta function into Eq. \eqref{eq:AverageOverReplicas} to enforce the constraint \eqref{eq:overlap}  we get
\begin{equation}
\begin{aligned}
&\int\left(\prod_{i=1}^{N}\prod_{a=1}^{n}\ \frac{\mathrm{d}\phi_i^a}{\sqrt{2\pi}}\right)\exp\left[-\frac{1}{2}\sum_{i=1}^{N}\sum_{a=1}^{n}\mathcal{E}\ (\phi_i^a)^2+
\frac{1}{4N}\sum_{i,k=1}^{N}\sum_{a,b=1}^{n}\phi_i^a\phi_k^a\phi_i^b\phi_k^b\right]=\\
&=\int\left(\prod_{a,b=1}^{n}\mathrm{d}q_{ab}\right)
\exp\left[-\frac{1}{2}\mathcal{E}N\sum_{a=1}^{n}q_{aa} + \frac{1}{4}N\sum_{a,b=1}^nq_{ab}^2\right]
\int\left(\prod_{i=1}^{N}\prod_{a=1}^{n}\ \frac{\mathrm{d}\phi_i^a}{\sqrt{2\pi}}\right)\prod_{a,b}\delta\left(Nq_{ab}-\sum_{i=1}^{N}\phi_i^a\phi_i^b\right)\ .\label{eq:IntroOverlap}
\end{aligned}
\end{equation}
In the second line of Eq. \eqref{eq:IntroOverlap} we have omitted a prefactor independent from $q_{ab}$. In the following we will neglect any multiplicative constant as well. 
Let us consider the last integral in Eq. \eqref{eq:IntroOverlap}. Using the integral representation of the $\delta$-function, we find
\begin{equation}
\begin{aligned}
&\int\left(\prod_{i=1}^{N}\prod_{a=1}^{n}\ \frac{\mathrm{d}\phi_i^a}{\sqrt{2\pi}}\right)\prod_{a,b}\delta\left(Nq_{ab}-\sum_{i=1}^{N}\phi_i^a\phi_i^b\right)=\\
&\int\left(\prod_{a,b=1}^n\mathrm{d}\psi_{ab}\right)\int\left(\prod_{i=1}^{N}\prod_{a=1}^{n}\ \frac{\mathrm{d}\phi_i^a}{\sqrt{2\pi}}\right)\exp\left(\frac{N}{2}\sum_{a,b=1}^nq_{ab}\psi_{ab}-\frac{1}{2}\sum_{i=1}^N\sum_{a,b=1}^n\phi_i^a\psi_{ab}\phi_i^b\right)=\\
&\int\left(\prod_{a,b=1}^n\mathrm{d}\psi_{ab}\ \mathrm{e}^{\frac{N}{2}q_{ab}\psi_{ab}} \right)
\left[\int\left(\prod_{a=1}^{n}\ \frac{\mathrm{d}\phi^a}{\sqrt{2\pi}}\right)\exp\left(-\frac{1}{2}\sum_{a,b=1}^n\phi^a\psi_{ab}\phi^b\right)\right]^N=\\
&\int\left(\prod_{a,b=1}^n\mathrm{d}\psi_{ab}\ \right)\exp\left(\frac{N}{2}\sum_{a,b=1}^n q_{ab}\psi_{ab}-\frac{N}{2}\mathrm{Tr}\log(\hat{\psi})\right)\ ,
\end{aligned}
\end{equation}
where $\hat{\psi}$ is the matrix with entries $\psi_{ab}$, and the integral over $\psi_{ab}$ is performed on the imaginary axis. 

Since we are interested in the limit $N\to\infty$, we can evaluate the integral using the steepest descent method and we get the
following saddle point equations:
\begin{equation}
q_{ab}=(\hat{\psi}^{-1})_{ab}\ .
\end{equation}
Evaluating the integral at the saddle point, we finally obtain (always discarding terms which do not depend on $\hat{q}$) 
\begin{equation}
\int\left(\prod_{i=1}^{N}\prod_{a=1}^{n}\ \frac{\mathrm{d}\phi_i^a}{\sqrt{2\pi}}\right)\prod_{a,b}\delta\left(Nq_{ab}-\sum_{i=1}^{N}\phi_i^a\phi_i^b\right)\ =\ 
\exp\left(\frac{N}{2}\mathrm{Tr}\log(\hat{q})\right)\ .
\label{eq:Jacobian}
\end{equation}

Coming back to Eq. \eqref{eq:IntroOverlap}, and using the result of Eq. \eqref{eq:Jacobian}, we find
\begin{equation}
\overline{\left(\frac{1}{\det(\mathcal{E}\mathbb{I}-\hat{J})}\right)^{n/2}}\ =\
 \int\Bigg(\prod_{a,b=1}^{n}\mathrm{d}q_{ab}\Bigg)
\exp\left(-\frac{1}{2}Nf_n[\hat{q}]\right)\ ,
\label{eq:detSP}
\end{equation} 
where the function $f_n[\hat{q}]$ is given by
\begin{equation}
f_n[\hat{q}]\ =\ \mathcal{E}\sum_{a=1}^{n}\ q_{aa}\  -\  \frac{1}{2}\sum_{a,b=1}^nq_{ab}^2\ -\ 
\mathrm{Tr}\log(\hat{q})\ .
\end{equation}

The integral on the r.h.s. of Eq. \eqref{eq:detSP} can be still evaluated via the saddle point method in the limit $N\to\infty$.  The key issue is to identify the correct saddle point. Usually, if the free energy is invariant under the action of a group, the first choice is a saddle point that is invariant under the group.
 
The most evident group of symmetry of the function $f_n[\hat{q}]$ is the permutation group of $n$ elements \footnote{In this case, but not in the spin glass case the symmetry group, is larger, i.e. it is the $O(n)$ group, because $f_n[\hat{q}]$ depends only on the trace of $\hat{q}$.}, i.d. we can permute simultaneously the rows and the columns of the matrix $q$ (this point will be discussed at lengthy later in these notes).

We make now the \textit{replica symmetric} ansatz on the form of the saddle point solution, i.e., we assume that the \textit{free energy} $f_n[\hat{q}]$ has a minimum on the subspace of matrices $\hat{q}$ that are invariant under the permutation group. These matrices are of the form \footnote{The $O(n)$ symmetry would imply $p=0$.}:
\begin{equation}
q_{ab}=q\delta_{ab}+p(1-\delta_{ab})\ .
\end{equation}
The saddle point equations then become 
\begin{equation}
\begin{aligned}
&\frac{\partial f_n(q,p)}{\partial q}=n\left[\mathcal{E}-q-\frac{1}{q-p}+\frac{p}{(q-p)^2}\right]=0\ \\
&\frac{\partial f_n(q,p)}{\partial p}=n\left[p-\frac{p}{(q-p)^2}\right]=0.
\end{aligned}
\end{equation}
The meaningful solution is given by 
 \begin{equation}
\begin{aligned}
&q^*=\frac{\mathcal{E}-\sqrt{\mathcal{E}^2-4}}{2}\ .\\
&p^*=0\ .
\end{aligned}
\end{equation}
Evaluating the integrand of Eq. \eqref{eq:detSP} at the saddle point, we obtain the following result:
\begin{equation}
\overline{\left(\frac{1}{\det(\mathcal{E}\mathbb{I}-\hat{J})}\right)^{n/2}}=
\exp\left(-\frac{1}{2}Nf_n(q^*,p^*)\right)\ .\label{eq:detAve}
\end{equation} 
The function $f_n(q^*,p^*)$, in the small $n$ limit, can be written as $f_n(q^*,p^*)=nf(q^*,p^*)+O(n^2)$, where the function $f(q,p)$ is given by
\begin{equation}
f(q,p)\ =\ \mathcal{E}q - \frac{1}{2}(q^2-p^2)-\log(q-p)-\frac{p}{q-p}
\end{equation}

From expression \eqref{eq:detAve}, using Eqs. \eqref{eq:TraceResolvent} and \eqref{eq:Introreplica}, we can easily compute the trace of the resolvent in the limit $N\to\infty$:
\begin{equation}
\begin{aligned}
&\lim_{N\to\infty}\frac{1}{N}\ \overline{\mathrm{Tr}\left[\hat{R}(\mathcal{E})\right]}=
-2\ \frac{\mathrm{d}}{\mathrm{d}\mathcal{E}}\ 
\lim_{n\to 0}\ \frac{\mathrm{d}}{\mathrm{d}n}\
\left[\lim_{N\to\infty}
\frac{1}{N}\  \overline{\left(\frac{1}{\det(\mathcal{E}\mathbb{I}-\hat{J})}\right)^{n/2}}\ \right] 
=\frac{\mathrm{d}}{\mathrm{d}\mathcal{E}}\ f(q^*,p^*)\ .\label{eq:DerFreeEn}
\end{aligned}
\end{equation}
The l.h.s. of the Eq. \eqref{eq:DerFreeEn} is given by
\begin{equation}
\frac{\mathrm{d}}{\mathrm{d}\mathcal{E}}\ f(q^*,p^*)=
\frac{\partial}{\partial\mathcal{E}}\ f(q^*,p^*)=q^*\ ,
\end{equation}
because of the stationarity of $f(q,p)$ with respect to variation in $q$ and $p$. So, we finally find
\begin{equation}
\lim_{N\to\infty}\ \frac{1}{N}\ \overline{\mathrm{Tr}\left[\hat{R}(\mathcal{E})\right]}\ =\ \frac{\mathcal{E}-\sqrt{\mathcal{E}^2-4}}{2}\ .
\end{equation}
We can now follow exactly the same lines after Eq. \eqref{eq:AsymptResol} to recover the Wigner distribution $\rho(\mathcal{E})$, given by Eq. \eqref{eq:Wigner}.

\section{The Sherrington-Kirkpatrick model}

Mean-field spin-glass models are spin systems with random interactions ($J$) in which there is no notion of space or distance 
and each degree of freedom interacts with all the others. 

Here we consider the paradigmatic example of the Sherrington-Kirkpatrick model, defined by the following Hamiltonian
\begin{equation}
{\mathcal H}[\sigma]=-\sum_{\langle i,j\rangle}^N J_{ij} \sigma_i \sigma_j - h\sum_i^N \sigma_i \, ,
\end{equation}
where $\sigma_i\in\lbrace -1,1\rbrace$, $h$ is the uniform magnetic field and the couplings are random variables 
extracted from the distribution
\begin{equation}
P(J)=\sqrt{\frac{N}{2\pi}} \exp\left({-\frac{NJ^2}{2}}\right) \, .
\end{equation}

The randomicity of the interaction is called {\it quenched disorder}, where the word {\it quenched} underlines the fact 
that the thermodynamic properties have to be computed at a fixed instance of the disorder. When the size 
of the system goes to infinity, thanks to the {\it self-averaging} property, the free energy of the single sample (specific 
realization of the disorder) is given by the average over the disorder of the $J$-dependent free energy, namely we have 
\begin{equation}
\lim_{N\rightarrow \infty} f^{(N)}_{J} = f = \lim_{N\rightarrow \infty} \overline{f^{(N)}_{J}}\, ,
\end{equation} 
where the overbar denotes the average over the disorder according to its distribution (Gaussian or Bimodal for example). The previous relation is satisfied with probability one: there are sequences of the $J$'s such that the previous result does not hold (e.g. all $J^N_{i,k}=1$) but these sequences have zero measure.

Therefore the solution of the statics of such a system in presence of quenched disorder requires the computation 
of the average of the logarithm of the (sample dependent) partition function 
\begin{equation}
Z_J=\sum_{\lbrace \sigma \rbrace} \exp\left({-\beta {\mathcal H}[\sigma]}\right)=\sum_{\lbrace \sigma \rbrace} \exp\left({+\beta\sum_{\langle i,j\rangle} J_{ij} \sigma_i \sigma_j +\beta h\sum_i \sigma_i}\right)\, ,
\end{equation}
which unfortunately cannot be done analytically. 

The so-called {\it replica trick} allows to overcome this difficulty, in fact the following identity holds 
\begin{equation}
\lim_{n\to 0} \frac{\overline{Z^n}-1}{n}=\overline{\log Z}
\end{equation}
or, equivalently
\begin{equation}
\lim_{n\to 0} \frac{\log(\overline{Z^n})}{n}=\overline{\log Z} \,.
\end{equation}
Computing the average of the $n$-th power of the partition function is a much easier task than 
computing the average of its logarithm. One actually computes the quantity for an integer $n$ and than 
takes a (a posteriori) harmless analytic continuation to real $n$ in order to reach the limit zero:
\begin{equation}
\overline{Z_J^n}=\int \left( \prod_{i<j} \, dJ_{ij} \, P(J_{ij}) \right) \sum_{\lbrace \underline{\sigma} \rbrace} \exp\left({\beta\sum_a\sum_{\langle i,j\rangle} J_{ij} \sigma^a_i \sigma^a_j +\beta h \sum_a \sum_i \sigma^a_i}\right)\,.
\end{equation}
When $n$ is integer we actually deal with the partition function of $n$ non-interacting copies (replicas) of the 
system of interacting spins living in the same realization of the disorder.  At the end of the computation, after the disorder average is 
taken, we will end up with an effective action that must be {\sl minimized} with respect to an order parameter that is 
an $n\times n$ symmetric matrix. Let us now proceed with the computation of the replicated partition function. 

In the following we will largely make use of the identities:
\begin{eqnarray}
&&\int \, dx \, \exp\left({-Ax^2 + B x }\right)=\sqrt{\frac{\pi}{A}} \exp\left({\frac{B^2}{4A}}\right)
\\
\nonumber
&&\int \, d^Mx \, \exp\left({-\sum_{i,k=1,M}A_{i,k}x_i x_k + \sum_{i,M}B_i x_i }\right)=\sqrt{\frac{\pi}{\mathrm{det}(A)}} \exp\left(\frac14 \sum_{i,k=1,M}\left(\hat{A}^{-1}\right)_{i,k}B_i B_k \right) \,,
\label{hubbard}
\end{eqnarray}
that is a Gaussian integration if considered from left to right or a so-called {\it Hubbard-Stratonovich} (H-S) transformation if 
considered from right to left. 

If we consider the formula (\ref{hubbard}) we can easily get rid of the disorder average obtaining
\begin{equation}
\overline{Z_J^n}=\sum_{\lbrace \underline{\sigma} \rbrace} \exp\left({\beta h \sum_i \sum_a \sigma^a_i + \frac{\beta^2}{2N} \sum_{i<j} \sum_{a,b} \sigma^a_i \sigma^b_i \sigma^a_j \sigma^b_j}\right) .
\end{equation}

Note that, performing the Gaussian integration we have introduced an ``interaction'' between replicas which, therefore, are no longer 
independent. The interaction term can be rewritten in the following way:
\begin{equation}
\sum_{i<j} \sum_{a,b} \sigma^a_i \sigma^b_i \sigma^a_j \sigma^b_j= N^2 \sum_{a<b} \left(  \frac{1}{N} \sum_i \sigma^a_i \sigma^b_i\right)^2 +\frac{N^2 n - Nn^2}{2}\,.
\end{equation}
The replicated partition function reads
\begin{equation}
\overline{Z_J^n}=\sum_{\lbrace \underline{\sigma} \rbrace} \exp\left({\beta h \sum_i \sum_a \sigma^a_i}\right) \exp\left({\frac{\beta^2}{4} (Nn-n^2)}\right) \prod_{a<b} \exp\left({\frac{\beta^2}{2N} (\sum_{i} \sigma^a_i \sigma^b_i)^2}\right)\,. 
\end{equation}
At this point we perform a H-S transformation (\ref{hubbard}) with
\begin{equation}
4A=2N \beta^2\,\,\, , \,\,\, B^2=\left( \beta^2 \sum_i \sigma^a_i \sigma^b_i \right)^2 \,\,\, , \,\,\, C=0 \,,
\end{equation}
so that the partition function becomes
\begin{eqnarray}
&&\overline{Z_J^n}=\sum_{\lbrace \underline{\sigma} \rbrace} \exp\left({\beta h \sum_i \sum_a \sigma^a_i}+{\frac{\beta^2}{4} (Nn-n^2)}\right) \left( \frac{2\pi \beta^2}{N} \right)^{\frac{n(n-1)}{2}} \cdot \\
\nonumber
&&\prod_{a<b} \int \, dQ_{ab} \, \exp\left({-\frac{N}{2} \beta^2 Q^2_{ab}+\beta^2 \sum_i \sigma^a_i \sigma^b_i \, Q_{ab}} \right)\,,
\end{eqnarray} 
where finally replicas are coupled and spins inside one replica have been decoupled at the price of introducing 
an integration over the matrix $Q_{ab}$. This formula can be also written  more explicitly as
\begin{eqnarray}
&&\overline{Z_J^n}= \exp\left({\frac{\beta^2}{4} (Nn-n^2)}\right) \left( \frac{2\pi \beta^2}{N} \right)^{\frac{n(n-1)}{2}} \cdot \\
\nonumber
&&\int \, dQ \, \exp\left({-\frac{N}{2} \beta^2 \sum_{a<b} Q^2_{ab}}\right) \left(\sum_{\lbrace \underline{\sigma} \rbrace} \exp\left({\beta h \sum_a \sigma^a +\beta^2 \sum_{a<b} \sigma^a \sigma^b \, Q_{ab}} \right)^N\right)\,.
\label{rep}
\end{eqnarray}
We are now interested in computing the free energy
\begin{equation}
f(\beta, h)= \lim_{n\to 0} \lim_{N\to \infty} \left( -\frac{1}{\beta N n} \ln \overline{Z_J^n} \right)\,.
\end{equation}

Note that, in principle, the two limits should have been taken in the opposite order, nevertheless the calculation would be impossible 
with the right order of the limits. For this reason we ``blindly'' exchange the order assuming that this operation is harmless. 
Since $N\to \infty$, we will make a saddle point calculation, therefore in the expression of the replicated partition function (\ref{rep}) 
we can discard all the terms that are not exponential in $N$. In addition we can retain only terms exponential in $n$, dropping terms 
that are exponential in $n^2$ since, in the final limit $n\to 0$ the latter would vanish. 
Given the above comments, we can express the replicated partition function as follows
\begin{equation}
\overline{Z_J^n} \propto \int dQ \, \exp\left({-N{\mathcal S}[Q,h]}\right)
\end{equation}
with
\begin{equation}
{\mathcal S}[Q,h]=-\frac{\beta^2 n}{4}  + \frac{\beta^2}{2} \sum_{a<b} Q^2_{ab} - {\mathcal W}[Q]
\label{action}
\end{equation}
and
\begin{equation}
{\mathcal W}[Q]=\ln \sum_{\lbrace \underline{\sigma} \rbrace} \exp\left({\beta h \sum_a \sigma^a +\beta^2 \sum_{a<b} \sigma^a \sigma^b \, Q_{ab}}\right)
\end{equation}

Taking the saddle point w.r.t. the order parameter $Q$ we obtain that the free energy can be written as
\begin{equation}
f(\beta, h)= \lim_{n\to 0} \frac{1}{\beta n}\,\, \mathrm{extr}_{Q} \,\,{\mathcal S}[Q,h] \,,
\end{equation}
where we used {\it extremal point} instead of {\it minimum} because, in the $n\to 0$ limit, minima become 
maxima, but the discussion of this topic is beyond the scope of these lecture notes (see \cite{MPV,Guerra} and references therein).  At the end of the day one finds that a good saddle point must satisfy the condition that the Hessian matrix,
\begin{equation}
M_{(ab),(cd)}=\frac{\partial^2 {W}}{\partial Q_{a,b}\partial Q_{c,d}} \,,
\end{equation}
has non-negative eigenvalues.  In a {\sl bona fide} Hilbert space this condition  is a prerequisite for having a minimum.

The minimization of the ``effective action'' ${\mathcal S}$ leads to a self-consistency equation for the order parameter, namely 
\begin{equation}
Q_{ab}=\langle \langle s^a s^b \rangle \rangle \,,
\end{equation}
where $\langle\langle \cdots \rangle\rangle$ stands for the expectation value taken w.r.t. the probability measure
\begin{equation}
\mu[\underline{\sigma}]=\frac{\exp\left({\beta h \sum_a \sigma^a +\beta^2 \sum_{a<b} \sigma^a \sigma^b \, Q_{ab}}\right)}{\sum_{\lbrace \underline{\tau} \rbrace} \exp\left({\beta h \sum_a \tau^a +\beta^2 \sum_{a<b} \tau^a \tau^b \, Q_{ab}}\right)}\,.
\label{meas}
\end{equation}

It should be evident at this point that performing the minimization of the effective action with a generic $n\times n$ symmetric matrix 
is an impossible task to accomplish. We need to parametrize the matrix in such a way that makes explicit the $n$ dependence and 
makes it possible to take the $n\to 0$ limit. Note also that one important requirement on the parametrization we choose is that 
the effective action contains terms that goes to zero {\it at least} linear in $n$, otherwise the free energy would diverge in the zero-replicas limit. 
In the next Section we show the simplest parametrization we can choose and the results to which it leads.

\subsection{The replica symmetric solution} 

Let us start with an elementary example that introduces us to the more difficult task we are trying to accomplish. 
Consider a function of two variables $g$, such that
\begin{equation}
g(x,y)=g(y,x)
\end{equation}
This function is symmetric under the exchange of the two arguments. In this simple case it is easy to understand the consequences 
of such a symmetry in the process of minimizing (or in general extremizing) the function $g$. 
When looking for a point of minimum $(x^*,y^*)$ of the above function we have clearly two possibilities:

\begin{itemize}
\item The first possibility is that $x^*=y^*$, which means that the extremal point is {\it invariant} under the action 
of the symmetry of the problem, namely the exchange of the two variables. In order to find such an extremal point we could 
easily restrict to the line $x=y$ and minimize $g$ in this one-dimensional subspace. 
\item The second possibility is that $x^*\neq y^*$. If this is the case, both $(x^*,y^*)$ and $(y^*,x^*)$ are necessarily 
minimum points. If this happens, the symmetry is said to be {\it broken} and the action of the symmetry group of the 
problem transforms one extremal point into the other and viceversa. The way in which the symmetry must be broken, as we will see, 
is a completely non-trivial issue in many situations.
\end{itemize}

The problem we are facing is, despite its more intricate nature, absolutely analogous. 
In fact the effective action (\ref{action}) is symmetric under the permutation group over replicas, which means that 
\begin{equation}
{\mathcal S}[\lbrace Q_{ab}\rbrace]={\mathcal S}[\lbrace Q_{\pi(a) \pi(b)}\rbrace] \,,
\end{equation}
where $\pi$ is an arbitrary permutation of replica indices.
As in the example above we want to start with the simplest (and historically first) {\it Ansatz} possible: we assume 
that the solution is invariant under the symmetry group of the action, namely, if $Q^*$ is a minimum point, we have 
\begin{equation}
\lbrace Q^*_{ab} \rbrace = \lbrace Q^*_{\pi(a) \pi(b)} \rbrace \,,
\end{equation}
{\it i.e.} the matrix does not changing under an arbitrary renumbering of replicas. 
It is easily shown that the only invariant matrix is the so-called replica-symmetric (RS) one
\begin{eqnarray}
&&Q_{aa}=0\\
\nonumber
&&Q_{ab}=q \,\,\,\,\,\,\,\,\, a\neq b \,,
\end{eqnarray}
where all the diagonal elements are zero and all the off-diagonal elements take the same value $q$. 

We can plug this {\it Ansatz} into the effective action (\ref{action}) obtaining
\begin{eqnarray}
\nonumber
{\mathcal S}[q,h]&&=-\frac{n \beta^2}{4}-\frac{n \beta^2 q^2}{4} -\ln \sum_{\lbrace \underline{\sigma} \rbrace} \exp\left(\beta h \sum_a \sigma^a +\frac{\beta^2}{2} q \sum_{a\neq b} \sigma^a \sigma^b \right) \\
\nonumber
&&=-\frac{n \beta^2}{4}-\frac{n \beta^2 q^2}{4} -\ln \sum_{\lbrace \underline{\sigma} \rbrace} \exp\left(\beta h \sum_a \sigma^a  \right) \exp\left(-\frac{n\beta^2 q}{2}\right)\, \left(\frac{\beta^2}{2} q \left(\sum_{a} \sigma^a\right)^2\right) \\ 
\nonumber
&&=-\frac{n \beta^2}{4}-\frac{n \beta^2 q^2}{4} -\ln \sum_{\lbrace \underline{\sigma} \rbrace} \exp\left(\beta h \sum_a \sigma^a  \right) \exp\left(-\frac{n\beta^2 q}{2}\right)\, \int \frac{dz}{\sqrt{2\pi}} \, \exp\left(-\frac{z^2}{2} + \beta \sqrt{q} z \sum_{a} \sigma^a\right) \\
&&=-\frac{n \beta^2}{4}-\frac{n \beta^2 q^2}{4} +\frac{n \beta^2 q}{2} -\ln  \int \frac{dz}{\sqrt{2\pi}} \, \exp\left(-\frac{z^2}{2}\right)  \left( 2 \cosh(\beta h+ \beta \sqrt{q} z) \right)^n\\
\nonumber
&&=-\frac{n \beta^2}{4}-\frac{n \beta^2 q^2}{4} +\frac{n \beta^2 q}{2} -\ln  \int \frac{dz}{\sqrt{2\pi}} \, \exp\left(-\frac{z^2}{2}\right)  \exp\left(n \ln \left( 2 \cosh(\beta h+ \beta \sqrt{q} z) \right)\right)\\
\nonumber
&&=-\frac{n \beta^2}{4}-\frac{n \beta^2 q^2}{4} +\frac{n \beta^2 q}{2} -\ln \left( 1+ n \,\int \frac{dz}{\sqrt{2\pi}} \, \exp\left(-\frac{z^2}{2}\right)   \ln \left( 2 \cosh(\beta h+ \beta \sqrt{q} z) \right)\right) \\
\nonumber
&&=-\frac{n \beta^2}{4}-\frac{n \beta^2 q^2}{4} +\frac{n \beta^2 q}{2} - n \,\int \frac{dz}{\sqrt{2\pi}} \, \exp\left(-\frac{z^2}{2}\right)   \ln \left( 2 \cosh(\beta h+ \beta \sqrt{q} z) \right)
\label{steps}
\end{eqnarray}
where, again, at the third line we have introduced a H-S transformation and all the steps are valid only in the $n\to 0$ regime. 
Dividing by $\beta$ and $n$ and taking the zero-replicas limit, we obtain the Gibbs free energy to be minimized w.r.t. $q$. 

The final  Gibbs free energy reads
\begin{equation}
\tilde{f}(q,h)=-\frac{\beta}{4}(1-q)^2-\frac{1}{\beta}\int \frac{dz}{\sqrt{2\pi}} \, \exp\left({-\frac{z^2}{2}} \right)  \ln \left( 2 \cosh(\beta h+ \beta \sqrt{q} z) \right)  \,.
\end{equation}
Taking the stationarity condition we obtain
\begin{equation}
\frac{\partial \tilde{f}(q,h)}{\partial q}= \frac{\beta}{2}(1-q) - \int \frac{dz}{\sqrt{2\pi}} \, \exp\left({-\frac{z^2}{2}}  \right)  \tanh(\beta h+ \beta \sqrt{q} z) \frac{z}{2\sqrt{q}}=0  \,.
\end{equation}

With one integration by parts and some simple algebraic manipulation one obtains the following canonical form for the 
self-consistency equation.
\begin{equation}
q=\int \frac{dz}{\sqrt{2\pi}} \, \exp\left({-\frac{z^2}{2}} \right)   \tanh^2(\beta h+ \beta \sqrt{q} z)  \,.
\label{selfRS}
\end{equation}

At zero magnetic field, Eq. (\ref{selfRS}) has a unique solution $q=0$ for $\beta < \beta_c=1$, while for $\beta > \beta_c$ another (physical) solution appears with $q\neq 0$. Therefore there is a phase transition at $h=0$ and $\beta=1$ while there is no phase transition 
for finite $h$. 
Despite everything seems to be consistent, an accurate computation shows that the entropy becomes negative at low temperature (in a discrete system this is forbidden by definition), 
in particular the zero temperature entropy is $S(0)=-\frac{1}{2\pi}\approx -0.17$. 
This result is an evidence that our assumption of Replica Symmetry is wrong at low temperatures, therefore in the next section 
we perform the stability analysis of the RS solution in some detail.

\subsection{Stability}
The stability of the replica symmetric solution was first studied in \cite{dAT}.

In order to analyze the stability of the solution in the full replica space, we consider the Hessian, which is an $\frac{n(n-1)}{2} \times \frac{n(n-1)}{2}$ 
symmetric matrix of the form
\begin{eqnarray}
\nonumber
M_{(ab),(cd)}&&=\frac{\partial^2 W}{\partial Q_{ab} \partial Q_{cd}} \\
\nonumber
&&=\frac{\partial}{\partial Q_{cd}} \left[ \beta^2 Q_{ab} - \beta^2 \frac{\sum_{\lbrace \underline{\sigma} \rbrace} \sigma^a \, \sigma^b \, \exp\left({\beta h \sum_f \sigma^f +\beta^2 \sum_{f<g} \sigma^f \sigma^g \, Q_{fg}}\right)}{\sum_{\lbrace \underline{\sigma} \rbrace} \exp\left({\beta h \sum_f \sigma^f +\beta^2 \sum_{f<g} \sigma^f \sigma^g \, Q_{fg} }\langle  \rangle\right)} \right]\\
&&=\frac{\partial}{\partial Q_{cd}} \left[ \beta^2 Q_{ab} - \beta^2 \langle \langle \sigma^a \sigma^b \rangle \rangle \right] \\
\nonumber
&&=\beta^2 \delta_{(ab),(cd)} - \beta^4 \left[ \langle\langle \sigma^a \sigma^b \sigma^c \sigma^d \rangle\rangle - \langle\langle \sigma^a \sigma^b \rangle\rangle \langle\langle \sigma^c\sigma^d \rangle\rangle \right] \,,
\end{eqnarray}
where, again, the double angular brackets stand for the average w.r.t. the measure (\ref{meas}). 

Since we are computing the Hessian in a RS point, it is easy to realize that it contains three kinds of elements, namely
\begin{equation}
M_{(ab),(cd)}= \begin{cases} 
M_1 \quad \quad \quad (ab)=(cd)\\ 
M_2 \quad \quad \quad a=c, b\neq d \,\,\, {\mathrm or}  \,\,\, a\neq c, b=d \\
M_3 \quad \quad \quad a\neq c ,  b\neq d  \,,
\end{cases}
\end{equation}
that we can compute explicitly through steps similar to the ones in (\ref{steps}), obtaining
\begin{eqnarray}
\nonumber
M_1 &&= \beta^2 - \beta^4 \left( 1-q^2 \right) \\
M_2 &&= -\beta^4 \left( q-q^2 \right)\\
\nonumber
M_3 &&= -\beta^4 \left( r-q^2 \right) \,,
\end{eqnarray}
where
\begin{equation}
r=\int \frac{dz}{\sqrt{2\pi}} \, \exp\left({-\frac{z^2}{2}}\right)    \tanh^4(\beta h+ \beta \sqrt{q} z) 
\end{equation}

Now we have to compute the eigenvalues of the stability matrix, the eigenvector equation reads
\begin{equation}
\sum_{(cd)} M_{(ab),(cd)} v_{(cd)}= m v_{(ab)} \,.
\end{equation}

The first straightforward thing to notice is that 
all the lines of the Hessian matrix have the same sum, this means that the subspace spanned by the vector with all equal components is 
a good eigenspace (we indicate this subspace of dimension $1$ as the ``longitudinal'' (scalar)). In fact  
\begin{equation}
\left( M_1 + 2(n-2) M_2 + \frac{(n-2)(n-3)}{2} M_3 \right) v = m_L v
\end{equation}
gives the equation for the first eigenvalue $m_L$ where $L$ stands for longitudinal. The reader can easily 
check that the subspace orthogonal to the longitudinal one can be divided in other two orthogonal eigenspaces called, respectively, 
the ``anomalous'' (vectorial) and ``replicon'' (tensorial). The subspaces are defined by
\begin{eqnarray}
\nonumber
&& \mathrm{LONGITUDINAL} \quad v_{(ab)}=v\\
&& \mathrm{ANOMALOUS} \quad \,\,\,\,\,\, \,\, v_{(ab)}=\frac{1}{2} (v_a + v_b) \quad ,\quad \sum_a v_a=0\\
\nonumber
&& \mathrm{REPLICON} \quad \quad \,\,\,\,\,\,\,\,\,\,  v_{(ab)} \quad , \quad \sum_b v_{ab}=0 \,.
\end{eqnarray}
Note that the dimensionality of the anomalous space is $n-1$ while the replicon has dimension $n(n-3)/2$ and, as it should be, the 
sum of the dimensionality of the three subspaces is $n(n-1)/2$. 
The eigenvalues are
\begin{eqnarray}
\nonumber
&&m_L= M_1 + 2(n-2) M_2 + \frac{(n-2)(n-3)}{2} M_3\\
&&m_A= M_1 + (n-4) M_2 - (n-3) M_3\\
\nonumber
&&m_R= M_1 - 2 M_2 + M_3 \,.
\end{eqnarray}
In the limit $n\to 0$
\begin{eqnarray}
\nonumber
&&m_L= M_1 -4 M_2 + 3 M_3\\
&&m_A= M_1 -4 M_2 +3 M_3\\
\nonumber
&&m_R= M_1 - 2 M_2 + M_3\,,
\end{eqnarray}
where we notice that the longitudinal and anomalous spaces are degenerate. The onset 
of the instability is given by the first eigenvalue that becomes zero that is (the reader can verify it) the replicon. 
An explicit computation gives
\begin{equation}
m_R= \beta^2 - \beta^4 \left[ - (1-q^2) + 2(q-q^2) - (r-q^2) \right] \,.
\end{equation}

Combining the equation $m_R=0$ with the self-consistency (\ref{selfRS}) we obtain the 
definition of the so-called de Almeida-Thouless (dAT) line, the line of instability of the RS solution in the $(\beta-h)$ plane, namely
\begin{eqnarray}
\nonumber
1&&=\beta^2 \int  {\mathcal D} z\,    \mathrm{sech}^4(\beta h+ \beta \sqrt{q} z) \\
q&&=\int  {\mathcal D} z\,    \tanh^2(\beta h+ \beta \sqrt{q} z) \,,
\end{eqnarray}
where
\begin{equation}
{\mathcal D} z=\frac{dz}{\sqrt{2\pi}} \, \exp\left({-\frac{z^2}{2}}\right) \,.
\end{equation}

On the left of the dAT line the replica symmetric Ansatz is not valid and replica symmetry must be broken. We will 
see in the next section the Parisi breaking scheme that has been recently proven to be the one that provides the correct solution 
to the SK model.

\subsection{Breaking replica symmetry}

\begin{figure}[t]
\centering
\includegraphics[width=0.8\textwidth]{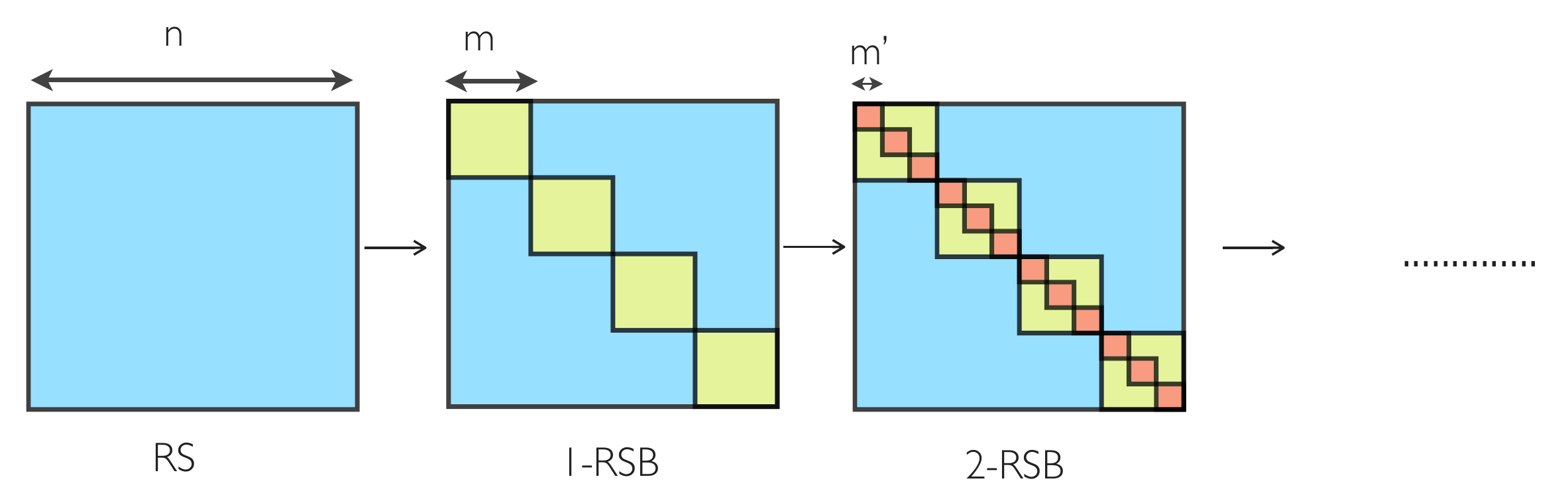}
\caption{A schematic representation of the Parisi Replica Symmetry Breaking scheme.}
\label{rsb}
\end{figure}

In this section we introduce in a formal way the iterative replica symmetry breaking scheme proposed by Parisi \cite{Parisi80a,Parisi80b}, while 
in the next section we will give a more physical interpretation of the RSB phenomenon. 
The Parisi solution is a sequence of Ansatz that approximate progressively better the true solution. At the end of the procedure 
the solution can be formulated in terms of a continuous function, defined in the limit of the infinite sequence.
We consider now, step by step, the way in which the solution is obtained. 
The first stage ($1$-RSB) goes as follows: the $n$ replicas are divided into $n/m$ groups of $m$ replicas, if two replicas 
$a$ and $b$ belong to the same group then the overlap matrix takes a value $q_1$, while if they belong to different groups it takes 
a value $q_0$. A compact way to express this structure is the following
\begin{equation}
Q_{ab}= \begin{cases} 
q_1 \quad \quad \quad I(a/m)=I(b/m)\\ 
q_0 \quad \quad \quad I(a/m)\neq I(b/m) \,,
\end{cases}
\end{equation}
where $I(z)$ is the integer part of $z$. A pictorial representation of the $1$-RSB scheme is given in Fig. \ref{rsb}. 
With computations similar to (\ref{steps}) (technically slightly more complicated) we can obtain the $1$-RSB free energy which is now a function of the 
three parameters $q_1$, $q_0$ and $m$. The Gibbs free energy reads 
\begin{eqnarray}
\nonumber
\tilde{f}(q_1,q_0,m)&&=-\frac{1}{4}\beta \left[ 1+mq_0^2+(1-m) q_1^2-2q_1 \right] -\frac{1}{m\beta} \int \frac{dz}{\sqrt{2\pi q_0}}\,\exp\left({-\frac{z^2}{2q_0}} \right)\\
&&\times \ln \left[\int \frac{dy}{\sqrt{2\pi (q_1-q_0)}} \, \exp\left({-\frac{y^2}{2(q_1-q_0)}}\right) \, (2 \cosh(\beta(z+y)))^m\right] \,.
\label{free1rsb}
\end{eqnarray}

In the process of taking the $n\to 0$ limit, we also have to promote the group size $m$, that was originally an integer 
between $1$ and $n$, to be a real number $m \in [0,1]$. Again, in the next section we will provide a more physical explanation 
for this fact.  
The resulting self-consistency equations for $q_1$, $q_0$ and $m$ can be obtained extremizing the free-energy (\ref{free1rsb}).
Notice that in the two extreme limits when $m=0$ and $m=1$ we actually recover a replica symmetric solution with, respectively, 
$q=q_0$ or $q=q_1$. 
Three comments are in order at this point:
\begin{itemize}
\item When minimizing the free-energy in the region of parameters where the replica symmetry is supposed to be broken (below 
the dAT line) $m$ is neither zero nor one as we expect. 
\item The zero temperature entropy is increased with respect to the replica symmetric one, in particular, one can obtain 
$S(0)\approx -0.01$
\item The computation of the Hessian eigenvalues shows that, close to the zero-field critical point, the replicon eigenvalue is less negative 
with respect to the RS case, more precisely $m_R=-C\tau^2/9$ compared to $m_R=-C\tau^2$ where $\tau=(1-T)$. 
\end{itemize}

All these facts suggest that, even though we have not found the correct solution, in some sense we have taken the right 
path.  
The procedure we have just described can be iterated and the single group of $m$ replicas can be divided into $m/m'$ groups of 
$m'$ replica. At this point, the matrix element between replicas belonging to the same subgroup will take a value $q_2$, between replicas belonging to the same group but not subgroup will take the value $q_1$ and, finally, $q_0$ if they belong to different groups. 
A pictorial representation of a $2$-RSB solution can be found in Fig. \ref{rsb}. 
We can now imagine to repeat the procedure $k$ times, obtaining the following definition for the matrix $Q$: 
\begin{equation}
Q_{ab}=q_i \,\,\,\,\,\,\,\,\,I(a/m_i)=(b/m_i) \,\,\,{\mathrm and} \,\,\, I(a/m_{i+1}=b/m_{i+1})\,\, :\,\,i=1,2,\cdots ,k+1 \,.
\end{equation}
In principle, for each of these scheme the computation of the free energy and the subsequent minimization can be repeated with respect to the parameters $q_1, \cdots, q_{k+1},m_1, \cdots, m_k$. The solution gets progressively better with the increasing number of steps, meaning that the zero temperature entropy becomes less and less negative and the instability at the transition weaker and weaker. This suggests the following ``natural'' step, that is sending the number of breakings to infinity.

In order to do this we define the following multi-step function in the interval $x\in [0,1]$:
\begin{equation}
q(x)=q_i \,\,\,\,\,\,\,\,\, 0\leq m_i<x<m_{i+1}\leq 1 :\,\,i=1,2,\cdots ,k \,.
\end{equation}
Taking the $k\rightarrow \infty$ limit, the free energy becomes a functional of $q(x)$ and the problem reduces to extremizing this 
functional w.r.t. all the possible functions $q$ with support in the unit interval, {\it i.e.} equating the functional derivative to zero 
\begin{equation}
\frac{\delta f[q]}{\delta q(x)}=0
\end{equation}
Generically the solution can be found numerically with a high degree of approximation (e.g. 8 decimal figures), but very close to the critical temperature 
an analytical treatment is possible expanding the free energy around $q(x)=0$. 

With some effort one can derive the algebra \cite{Parisi80b} that is associated to the matrices parametrized with the Parisi scheme (for any $n<1$ and $n\leq x\leq 1$), here we 
just present it briefly without derivation. 
A Parisi matrix is fully defined by its diagonal element and its off-diagonal function, namely 
\begin{equation}
A \rightarrow (\tilde{a},a(x)) \,.
\end{equation}
Now, consider two Parisi matrices and their Hadamard (dot) product $(A\cdot B)_{ab}=A_{ab} B_{ab}$, it 
is easy to prove that 
\begin{equation}
A\cdot B \rightarrow (\tilde{a} \tilde{b}, a(x)b(x))\,.
\end{equation}
Consider instead the usual matrix product $C=AB$, with some algebra it can be proved that 
\begin{equation}
\tilde{a}\tilde{b}-\langle ab \rangle \,,
\end{equation}
and 
\begin{equation}
c(x)=-n a(x)b(x)+[ \tilde{a}-\langle a \rangle]b(x)+[ \tilde{b}-\langle b \rangle]a(x) -\int_n^x\,dy\, [a(x)-a(y)][b(x)-b(y)]\,,
\end{equation}
where the compact notation 
\begin{equation}
\langle a \rangle = \int_n^1 \, dx\, a(x)
\end{equation}
has been used. 

Here we simply introduce the free energy in its full-RSB form (the derivation can be found in \cite{Parisi80b}).
The FRSB free energy reads 
\begin{equation}
f=-\frac{\beta}{4}\left[ 1+ \int_0^1 \, dx\, q^2(x) -2q(1)\right] - \frac{1}{\beta}\int \, {\mathrm D}u \, f_0(0,\sqrt{q(0)}u) \,, \label{freeRSB}
\end{equation} 
where $f_0$ is the solution to the non-linear antiparabolic equation 
\begin{equation}
\frac{\partial f_0(x,h)}{\partial x}=-\frac{1}{2} \frac{{\mathrm d} q}{{\mathrm d} x} \left[ \frac{\partial^2 f_0}{\partial h^2} + x \left( \frac{\partial f_0}{\partial h} \right)^2\right]\,,
\end{equation}
with the initial condition 
\begin{equation}
f_0(1,h)=\ln (2 \cosh  h)\ .
\end{equation} 
A nice collection of examples of solutions $q(x)$ in different situations can be found in Pagg. 40-43 of \cite{MPV}.

We remark that in spite of the numerical evidence that the solution of the stationary equation is unique, only recently it has been proved that the free energy in equation (\ref{freeRSB}) is convex and therefore its solution is unique \cite{CONV}.

\newpage

\section{Pure states, ultrametricity and stochastic stability in the RSB phase of the SK model}

The spin glass solution of the SK model is characterized by a spontaneous breaking of the replica symmetry, which corresponds to a non-trivial, i.e. non replica symmetric, form of the order parameter matrix $Q_{ab}$. As a consequence there are many solutions to the saddle point equations. This is due to the symmetry of the replica free energy function with respect to replica permutations. In other words, if there exists a particular solution for the matrix $Q_{ab}$ with a RSB structure, then any other matrix obtained via a renumbering of the replica indices in $Q_{ab}$ will be also a solution. It is also worth to note that the free energy barriers separating the corresponding RSB states must be infinite in the thermodynamic limit, since the mean field free energy is proportional to the volume of the system. Since all these states are stabilized in thermodynamic limit, we could call them the \textit{pure states} of the spin glass phase. On the other hand, the Gibbs state is obtained by summing up over all the pure states of the system, each pure state being taken with its statistical weight, which is defined by the value of the corresponding free energy. If we label the pure states with an index $\alpha$, we can define the corresponding thermodynamic weights $w_{\alpha}$ as \cite{MPVetal}\cite{Dotsenko} :
\begin{equation}
w_{\alpha}=\frac{\mathrm{e}^{-\beta F_{\alpha}}}{\sum_{\alpha}\mathrm{e}^{-\beta F_{\alpha}}}\ .
\end{equation} 
By definition, a pure state is a state in which the clustering property is satisfied, i.e., all correlation functions must factorize in the long distance limit. For example the two point correlation function must satisfy the aforementioned clustering property, which, in the case of the fully connected SK model, reads: $\langle\sigma_i\sigma_j\rangle_{\alpha}=\langle\sigma_i\rangle_{\alpha}\langle\sigma_j\rangle_{\alpha}$. The symbol $\langle\cdot\rangle_{\alpha}$ means that the Gibbs measure has to restricted to the pure state $\alpha$ only.

Let us now consider two different pure states of our system, labelled by two indices $\alpha$ and $\beta$, and let us define their mutual overlap $q_{\alpha\beta}$ as:
\begin{equation}
q_{\alpha\beta}\equiv\frac{1}{N}\sum_{i=1}^N\langle\sigma_i\rangle_{\alpha}\langle\sigma_i\rangle_{\beta}\ .
\end{equation}
To study the statistics of the whole set of possible overlaps $\{q_{\alpha\beta}\}$ it is useful to introduce the following probability distribution function
\begin{equation}
P_J(q)=\sum_{\alpha\beta}w_{\alpha}w_{\beta}\ \delta(q_{\alpha\beta}-q)\ .
\end{equation}
This distribution is sample-dependent, i.e., it depends on the specific realization of the random couplings $J_{ij}$. So, if we take the average of $P_J(q)$ over the disorder we get the \textit{physical} distribution function 
\begin{equation}
P(q)=\overline{P_J(q)}\ ,
\end{equation}
which gives the probability of finding a pair of pure states having the overlap equal to $q$. The distribution $P(q)$ can be considered as the physical order parameter. The fact that it is a function is a manifestation of the phenomenon that for the description of the spin-glass phase one needs an infinite number of order parameters.
Now we want to establish a contact between the physical order parameter $P(q)$ and the replica world. To this end let us first consider the following correlation function
\begin{equation}
q_J^{(k)}=\frac{1}{N^k}\sum_{i_1\dots i_k}\langle\sigma_{i_1}\dots\sigma_{i_k}\rangle^2\ .
\end{equation} 
Using the representation of the Gibbs average in terms of the pure states it is not difficult to show that 
\begin{equation}
q_J^{(k)}=\int\mathrm{d}q\ P_{J}(q)\ q^k\ .
\end{equation} 
Taking the average over the disorder one gets
\begin{equation}
q^{(k)}=\overline{q_J^{(k)}}=\int\mathrm{d}q\ P_{J}(q)\ q^k\ .
\end{equation} 
We see that the function $P(q)$, originally defined to describe the statistics of \textit{pure states}, can be practically calculated from the multipoint correlation functions in the \textit{Gibbs state}. Now it should be clear how to bridge replicas and physics. If we calculate the multipoint correlation functions using the replica approach, the connection of the physical order parameter with the Parisi RSB pattern would be established. The moment $q^{(k)}$ can be represented as follows:
\begin{equation}
q^{(k)}=\lim_{n\rightarrow 0}\ \left[\overline{\langle\sigma_i^a\sigma_i^b\rangle}\right]^k=\lim_{n\rightarrow 0}\ (Q_{ab})^k\ 
\end{equation} 
where $Q_{ab}=\overline{\langle\sigma_i^a\sigma_i^b\rangle}$ is the replica order parameter matrix, which is obtained from the saddle point equation for the replica free energy. Notice that, in the RSB phase, the entries of $Q_{ab}$ are not equivalent, and one has to sum over all the saddle point solutions for the matrix $Q_{ab}$ to perform the Gibbs average. Such solutions can be obtained from one of the RSB solutions by applying all possible permutations of rows and columns in $Q_{ab}$. The summation over all these permutations corresponds to the summation over the replica indeces $a$ and $b$ of the matrix $Q_{ab}$. As a consequence, the moment $q^{(k)}$ should be computed as follows:
\begin{equation}
q^{(k)}\ =\ \lim_{n\to0}\ \frac{1}{n(n-1)}\ \sum_{a\neq b}\ (Q_{ab})^k\ ,
\end{equation}
where the factor $n(n-1)$ is precisely the number of different replica permutations. From the previous equation one gets the following explicit expression for the distribution function $P(q)$:
\begin{equation}
P(q)\ =\ \lim_{n\to0}\ \frac{1}{n(n-1)}\ \sum_{a\neq b}\ \delta(Q_{ab}-q)\ . \label{eq:PqReplica}
\end{equation}
In the continuum $n\to0$ limit this result can be rewritten as
\begin{equation}
P(q)\ =\ \int_0^1\ {\mathrm d} x\ \delta(q(x)-q)\ .\label{eq:PqContinua}
\end{equation} 
Assuming that the function $q(x)$ is monotonous (which is the case for the RSB solution we are considering here), one can introduce the inverse function $x(q)$, and from Eq. \eqref{eq:PqContinua} one finds
\begin{equation}
P(q)\ =\ \frac{{\mathrm d} x(q)}{{\mathrm d} q} .
\end{equation}
This result is a key point: it defines the physical order parameter, i.e the distribution function $P(q)$, in terms of the formal saddle point function $q(x)$ \cite{GOLD}. By representing $x(q)$ in the following integral form 
\begin{equation}
x(q)\ =\ \int_0^q\ {\mathrm d} q'\ P(q') ,
\end{equation}
we can assign to this function a clear physical meaning: the function $x(q)$, inverse of $q(x)$, gives the probability of finding a pair of pure states having an overlap less than $q$.

\subsection{Fluctuations of $P_J(q)$}

Let us consider the function $P_J(q)$:
\begin{equation}
P_J(q)=\sum_{\alpha\beta} w_{\alpha} w_{\beta} \delta(q_{\alpha\beta} -q) .
\end{equation}
One reasonable problem to address is whether the probability $P_{J}(q)$ approaches a definite limit or fluctuates when the number $N$ of spins becomes infinite. To
 have an idea of why the latter case could be possible, let us consider, for a
 moment, an homogeneous ferromagnetic system ({\it i.e.} in absence of disorder). If we start from the high temperature phase and cool down below the Curie
temperature, the probabilities $w_+$ and $w_-$ of arriving to a state of magnetization $+m$ or $-m$, are well known to depend on the boundary conditions we have imposed on the system.

So, even in this simple case, $w_{\alpha}$, and therefore $P_J(q)$, are not good extensive quantities. The logical suspicion is that the same is true even in the more complicated spin glass case. Indeed, what happens is that $P_J(q)$ depends on the realization of the couplings, even in the thermodynamic limit. In other words, it is not a self averaging quantity, in the sense that, as far as $P_J(q)$ is concerned, an increasing size of the sample does not imply an average over all disorder configurations.

In order to quantify the fluctuations of $P_J(q)$, we will now compute $\overline{P_J(q_1)\ P_J(q_2)}-P(q_1)P(q_2)$. To this end we make use of the Laplace transform $g(y)$:
\begin{equation}
g(y)\ =\ \int{\mathrm d} q\ \overline{P_J(q)}\ \mathrm{e}^{yq}\ =\ \int{\mathrm d} q\ P(q)\ \mathrm{e}^{yq}\ .
\end{equation}
Using Eqs. (\ref{eq:PqReplica}) and (\ref{eq:PqContinua}) we can write $g(y)$ in the following way
\begin{equation}
g(y)\ =\ \frac{1}{n(n-1)}\ \sum_{a\neq b}\ \mathrm{e}^{yQ_{ab}}\ 
\xrightarrow[n \rightarrow 0]{}\ \int_0^1{\mathrm d} x\ \mathrm{e}^{yq(x)}\ .
\end{equation}
The computation of $\overline{P_J(q_1)\ P_J(q_2)}$ requires a slightly generalized Laplace transform $g(y_1,y_2)$:
\begin{equation}
g(y_1,y_2)\ =\ 
\int{\mathrm d} q_1\ {\mathrm d} q_2\ \overline{P_J(q_1)\ P_J(q_2)}\  \mathrm{e}^{y_1q_1+y_2q_2}\ =\ 
\int{\mathrm d} q_1\ {\mathrm d} q_2\ P(q_1,q_2)\   \mathrm{e}^{y_1q_1+y_2q_2}\ .
\end{equation}
$P(q_1,q_2)$ is the probability (averaged over $J$) to have overlaps $q_{\alpha_1\alpha_2}=q_1$ and $q_{\alpha_3\alpha_4}=q_2$ between $4$ pure states $\alpha_1,\alpha_2,\alpha_3,\alpha_4$. 
The function $g(y_1,y_2)$ can be computed by means of replicas using the following formula:
\begin{equation}
g(y_1,y_2)\ =\ \frac{1}{n(n-1)(n-2)(n-3)}\ \sum_{a\neq b\neq c\neq d=1}^n\ {\mathrm e}^{y_1Q_{ab}+y_2Q_{cd}}\ ,
\end{equation} 
where the sum is restricted to quadruplets of replicas $a,b,c,d$ which are all different. Using the RSB parametrization for $Q_{ab}$, and taking finally the limit $n\to0$, we get \cite{MPVetal}\cite{GhirlandaGuerra}:
\begin{equation}
\overline{P_J(q_1)\ P_J(q_2)}\ =\ \frac{1}{3}\ P(q_1)\ \delta(q_1-q_2)\ +\ \frac{2}{3}\ P(q_1)\ P(q_2)\ .
\end{equation}
This formula is the evidence that $P_J(q)$ fluctuates with $J$ even after the thermodynamic limit is taken.
There are infinite many relations of this kind as we will see in next subsection \cite{GhirlandaGuerra,SS1,SS2,PC}.

\subsection{Stochastic stability}
The principle of stochastic stability assumes that the distribution of the free energies of the various pure states is stable under independent random increments. 
The reason why we require such a principle is that the perturbations we use are random and they are not correlated with the original Hamiltonian. As a consequence they should change the free energies of the pure states by random amounts, leaving their distribution unaffected.

We present here the simplest non trivial case of stochastically stable system \cite{SS1}, where the overlaps have only two possible values: $q_0$ among different states and $q_1$ among the same state. In this case (which is usually called 1-step replica-symmetry breaking) we have only to specify the weight of each state, which is given by
\begin{equation}
w_\alpha\propto\mathrm{e}^{-\beta F_\alpha}\ .
\end{equation}

Let us consider the case where the $F$'s are independent random variables. The number of values of $F$ in the interval $[F,F+{\mathrm d} F]$ is given by
\begin{equation}
\rho(F){\mathrm d} F\ .
\end{equation}
Notice that as the number of states is infinite, the function $\rho(F)$ has a divergent integral. 

Let us now consider the effect of a perturbation of strenght $\epsilon$ on the free energy of a state, say $\alpha$. The unperturbed value of the free energy is denoted by $F_\alpha$. The new value of the free energy is given by $F'_\alpha=F_\alpha+\epsilon r_\alpha$, where $r_\alpha$ are identically distributed uncorrelated random variables. Stochastic stability implies that the distribution $\rho(F')$ is the same as $\rho(F)$. Expanding to second order in $\epsilon$, this condition leads to
\begin{equation}
\frac{{\mathrm d}\rho}{{\mathrm d} F} \propto \frac{{\mathrm d}^2\rho}{{\mathrm d} F^2}\ .
\end{equation}
The only physical solution (apart the trivial one $\rho(F)=0$, corresponding to non spin glass systems) is given by 
\begin{equation}
\rho(F)\propto\exp(\beta m F)\ ,\label{eq:rhoFromSS}
\end{equation} 
with an appropriate value of $m$. The parameter $m$ must satisfy the condition $m<1$, in order for the sum $\sum_\alpha\exp(-\beta F_\alpha)$ to be convergent. We see that  stochastic stability fixes the form of the distribution $\rho$, and hence connects the low and the high free energy parts of the function $\rho$. In other words, stochastic stability relates the property of the low lying configurations (that dominate the Gibbs measure) to those of the configurations much higher in free energy (that usually dominate the dynamics).
Accordingly, the requirement of stochastic stability gives also, nearly for free, some informations on the dynamics in the aging regime. In the dynamical evolution from an higher temperature initial state, the difference between the total free energy at time $t$ and the equilibrium value will be always of order $N$, with a prefactor going to zero when the time goes to infinity. So, one could argue that dynamics probes the behaviour of the function $\rho(F)$ at very large argument, and should not be related to the static properties, that, on the contrary, depend on the distribution $\rho$ for small values of the argument. However, stochastic stability forces the function $\rho(F)$ to be of the form \eqref{eq:rhoFromSS}, also in the range where $F$ is exstensive but small (let's say of order $\epsilon N$), and the previous objection can be discarded. 

The function $P(q)$, in the 1-step RSB case, is given by
\begin{equation}
P(q)=m\delta(q-q_0)+(1-m)\delta(q-q_1)\ .
\end{equation}
So we see that the same parameter $m$ enters both in the form of the function $P(q)$, which is dominated by the lowest values of $F$ (i.e. those producing the largest $w$'s), and in the form of the function $\rho(F)$ at large values of $F$.
This result is deeply related to the existence of only one family of stochastic stable systems with uncorrelated variables $F$.

In the case of full RSB the construction is more complex and it is described in details \cite{MPV,RC}.

\subsection{Ultrametricity}

The order parameter $P(q)$, far from being a parameter, indicates that the structure of the space of the spin glass pure states must be highly non trivial. However, the distribution function $P(q)$ of the pure states overlaps does not give enough information about this structure. To get insight into the topology of the space of pure states one needs to know the properties of the higher order correlations of the overlaps. So, let us consider any three pure states $\alpha_1$, $\alpha_2$, $\alpha_3$ and $P_J(q_1,q_2,q_3)$ the probability for them to have overlaps $q_1=q_{\alpha_2\alpha_3}$, $q_2=q_{\alpha_3\alpha_1}$, $q_3=q_{\alpha_1\alpha_2}$. In order to compute $P_J(q_1,q_2,q_3)$ one first consider the Laplace transform $g_J(y_1,y_2,y_3)$, and then uses replicas to calculate the average over the $J$'s \cite{MPVetal}\cite{MPV}: 
\begin{equation}
g(y_1,y_2,y_3)=\overline{g_J(y_1,y_2,y_3)}\ =\ \frac{1}{n(n-1)(n-2)}\ \sum_{a\neq b\neq c=1}^n\  
\mathrm{e}^{y_1Q_{ab}+y_2Q_{bc}+y_3Q_{ca}}\ ,
\end{equation}
where the sum must run over the triplets of replicas $a,b,c$ which are all different.
Taking the limit $n\to0$, after some algebra, one obtains the following result for $P(q_1,q_2,q_3)=\overline{P_J(q_1,q_2,q_3)}$:
\begin{equation}
\begin{aligned}
P(q_1,q_2,q_3)\ &=\ \frac{1}{2}\ P(q_1)x(q_1)\delta(q_1-q_2)\delta(q_1-q_3)\ +\  
\frac{1}{2}\ P(q_1)P(q_2)\theta(q_1-q_2)\delta(q_2-q_3) \\
&\ +  
\frac{1}{2}\ P(q_2)P(q_3)\theta(q_2-q_3)\delta(q_3-q_1)\ +\  
\frac{1}{2}\ P(q_3)P(q_1)\theta(q_3-q_1)\delta(q_1-q_2)\ .
\end{aligned}
\end{equation}
From this equation one can extract the following crucial property of the function $P(q_1,q_2,q_3)$. It is non-zero only in the following three cases: 
\begin{equation}
\begin{aligned}
&q_1\ =\ q_2 \leq q_3\ ,\\
&q_1\ =\ q_3 \leq q_2\ ,\\
&q_3\ =\ q_2 \leq q_1\ .
\end{aligned}
\end{equation}
In all other cases the function $P(q_1,q_2,q_3)$ is identically equal to zero. In words, this function is not zero only if at least two of the three overlaps are equal, and their common value is not bigger than the third one. It means that in the space of spin-glass pure states there exist no scalene triangles. More precisely, if we sample three configurations independently with respect to their common Gibbs distribution, and we average over the disorder, the distribution of the (Hamming) distances among them is supported, in the limit of very large system sizes, only on equilateral and isoscel triangles, with no contribution from scalene triangles.
The spaces having the above metric property are called \textit{ultrametric}. 

For a certain time it was believed that 
stochastic stability could be an independent property from ultrametricity, however there have been recently many papers suggesting the contrary, and this line of reasearch  culminated in the general proof of Panchenko that stochastic stability {\em does} imply ultrametricity \cite{PAT}.

At the end of the game we find the following surprising result.  In the case where the function $P_{J}(q)$ fluctuates when we change the parameters of the system, we can define its functional probability distribution ${\mathcal P}[P]$.  This functional order parameter is a description of the probability of $P_J(q)$ when we change the disorder \footnote{It may be possible that also for non-random systems we have a similar description where the average over the number of degrees of freedom plays the same role of the average over the disorder.}.  The functional probability distribution ${\mathcal P}[P]$ is an object that should have an infinite volume limit, i.e. ${\mathcal P}^\infty[P]$.

\newpage

\section{Thermodynamic limit in the SK model and exactness of the hierarchical RSB solution: rigorous results} 

\subsection{Existence of the thermodynamic limit}
The rigorous control of the infinite volume limit in the SK model can be very difficult, due to the effects of very large fluctuations produced by the external noise (the random couplings $J_{ij}$). Nonetheless Guerra and Toninelli \cite{GuerraToninelli} have introduced a very simple strategy for the control of the infinite volume limit. The main idea is to split a large system, made of $N$ spins, into two subsystems, made of $N_1$ and $N_2$ sites, respectively, where each subsystem is subject to some external noise, similar but independent from the noise acting on the large system. By a smooth interpolation between the system and the subsystems, one can show the \textit{subadditivity} of the quenched average of the free energy, with respect to the size of the system, and, therefore, obtain a complete control of the infinite volume limit. 

The Hamiltonian of the SK model, in a uniform external field of strength $h$, is given by 
\begin{equation}
H_{N}(\sigma,J,h)=-\frac{1}{\sqrt{N}}\sum_{j>i=1}^{N}J_{ij}\sigma_i\sigma_j-h\sum_{i=1}^{N}\sigma_i\ ,
\end{equation}
where the quenched disorder is given by the $N(N-1)/2$ independent and identical distributed random variables $J_{ij}$. Let us assume each $J_{ij}$ to be a Gaussian random variable with zero mean and unit variance.

For a given realization of the sample $J_{ij}$, the disorder dependent partition function $Z_N(\beta,J,h)$, at the inverse temperature $\beta$, is given by 
\begin{equation}
Z_N(\beta,J,h)=\sum_{\{\sigma\}}\exp\left[-\beta H_{N}(\sigma,J,h)\right]\ .
\end{equation} 
The quenched average of the free energy per spin $f_N(\beta,h)$ is
\begin{equation}
f_{N}(\beta,h)=-\frac{1}{\beta N\ }\mathbb{E}_J\log Z_N(\beta,J,h)\ .\label{eq:freeEn}
\end{equation}
 
Another important concept to recall is that of replicas. These are independent copies of the system, say $n$, characterized by the spin variables $\sigma_i^1,\sigma_i^2,\dots,\sigma_i^n$. All the replicas are subject to the same sample $J$ of the quenched disorder. The Boltzmann factor for the replicated system is given by
\begin{equation}
\exp\left[-\beta H_{N}(\sigma^1,J,h)-\beta H_{N}(\sigma^2,J,h)-\dots-\beta H_{N}(\sigma^n,J,h)\right]\ .\label{eq:BoltzRep}
\end{equation}
The overlaps between any two replicas $a,b$ are defined as
\begin{equation}
q_{ab}=\frac{1}{N}\sum_i\sigma_i^a\sigma_i^b\ ,\label{eq:overlap_ab}
\end{equation}
and they satisfy the bounds: $-1\leq q_{ab}\leq 1$. For a generic smooth function $F$ of the overlaps, we define the $\overline{\cdot}$ averages
\begin{equation}
\overline{ F(q_{12},q_{13},\dots)}=\mathbb{E}\langle F(q_{12},q_{13},\dots)\rangle_J\ ,
\end{equation}
where the thermal average $\langle\cdot\rangle_J$ is performed using the Boltzmann factor for the replicated system, Eq. \eqref{eq:BoltzRep}, and $\mathbb{E}$ is the average with respect to the quenched disorder  $J$.

Now we explain the main idea behind the Guerra-Toninelli method to control the infinite volume limit. Suppose to divide the $N$ spins into two blocks of $N_1$ and $N_2$ spins, with $N_1+N_2=N$, and define 
\begin{eqnarray}
Z_N(t)&&=\sum_{\{s\}}\exp\left(
\beta\sqrt{\frac{t}{N}}\sum_{j>i=1}^{N}J_{ij}\sigma_i\sigma_j
+\beta\sqrt{\frac{1-t}{N_1}}\sum_{j>i=1}^{N_1}J'_{ij}\sigma_i\sigma_j\right.\\
\nonumber
&&\left.+\beta\sqrt{\frac{1-t}{N_2}}\sum_{j>i=N_1+1}^{N_1}J''_{ij}\sigma_i\sigma_j
+\beta h\sum_{i=1}^{N}\sigma_i
\right)\ ,\label{eq:InterpolatingZ}
\end{eqnarray}
with $0\leq t\leq 1$. The quenched disorder is represented by the independent  families of random variables $J,J'$ and $J''$. The two subsystems are subjected to a different external noise, with respect to the original system, but, of course, the probability distributions are the same. The parameter $t$ allows us to interpolate between the original $N$ spin system at $t=1$ and a system composed of two non interacting parts at $t=0$. So we have
\begin{equation}
\begin{aligned}
Z_N(1)&=Z_N(\beta,h,J)\ ,\\
Z_N(0)&=Z_{N_1}(\beta,h,J')Z_{N_2}(\beta,h,J'')\ . \label{eq:boundaryCond}
\end{aligned}
\end{equation}
As a consequence, remembering Eq. \eqref{eq:freeEn}, we have
\begin{equation}
\begin{aligned}
-\beta^{-1}\ \mathbb{E}Z_N(1)&=Nf_N(\beta,h)\ ,\\
-\beta^{-1}\ \mathbb{E}Z_N(0)&=N_1f_{N_1}(\beta,h)+N_2f_{N_2}(\beta,h)\ .
\end{aligned}
\end{equation}
By taking the derivative of $-(N\beta)^{-1}\ \mathbb{E} Z_N(t)$ and performing a standard integration by parts on the Gaussian disorder, one obtains 
\begin{equation}
-\frac{1}{N\beta} \frac{{\mathrm d}}{{\mathrm d} t}  \mathbb{E} Z_N(t)=
\frac{\beta}{4} \left\langle  q_{12}^2 -\frac{N_1}{N}\left(q_{12}^{(1)}\right)^2-\frac{N_2}{N}\left(q_{12}^{(2)}\right)^2 \right\rangle \,,
\end{equation}
where 
\begin{equation}
\begin{aligned}
q_{12}^{(1)}&=\frac{1}{N_1}\sum_{i=1}^{N_1}\sigma_i^1\sigma_i^2\ ,\\
q_{12}^{(2)}&=\frac{1}{N_2}\sum_{i=N_1+1}^{N}\sigma_i^1\sigma_i^2\ .
\end{aligned}
\end{equation}

On the other hand $q_{12}$ is a linear combination of $q_{12}^{(1)}$ and $q_{12}^{(2)}$ with positive coefficients in the form 
\begin{equation}
q_{12}=\frac{N_1}{N}q_{12}^{(1)}+\frac{N_2}{N}q_{12}^{(2)}\ .
\end{equation}
Due to the convexity of the function $f :x\mapsto x^2$, we have the inequality
\begin{equation}
\left\langle q_{12}^2 -\frac{N_1}{N}\left(q_{12}^{(1)}\right)^2-\frac{N_2}{N}\left(q_{12}^{(2)}\right)^2\right\rangle \leq 0\ ,
\end{equation}

From this result we have a simple cascade of consequences. First of all the quenched average of the logarithm of the interpolating partition function, defined in \eqref{eq:InterpolatingZ}, is increasing in $t$:
\begin{equation}
\frac{1}{N}\frac{{\mathrm d}}{{\mathrm d} t}\ \mathbb{E}\log Z_N(t)\geq 0\ .
\end{equation}
By integrating in $t$ and recalling the boundary conditions \eqref{eq:boundaryCond}, we have 
\begin{equation}
Nf_{N}(\beta,h)\leq N_1f_{N_1}(\beta,h)+N_2f_{N_2}(\beta,h)\ .
\end{equation}
The previous inequality express the \textit{subadditivity} of  the quenched average of the free energy of the SK model. Since for every subadditive sequence $\{F_N\}_{N=1}^{\infty}$, the limit $\lim_{N\to\infty}\ F_N/N$ exists and is equal to $\inf_{N} F_N/N$ (Fekete's subadditive lemma), we have 

\begin{theorem}
\emph{(Guerra-Toninelli 1)}
\label{GT1}
The infinite volume limit for $f_{N}(\beta,h)$ does exists and equals its $\inf$
\begin{equation}
\lim_{N\to\infty}\ f_{N}(\beta, h)=\inf_{N}\  f_{N}(\beta, h)\equiv f(\beta,h)\ .\label{eq:GT1}
\end{equation}
\end{theorem}

After proving the existence of the thermodynamic limit for the quenched average, the result can be extended to prove that convergence holds for almost every disorder realization $J$. The following theorem is also due to Guerra and Toninelli:

\begin{theorem}
\emph{(Guerra-Toninelli 2)}
\label{GT2}
The infinite volume limit
\begin{equation}
\lim_{N\to\infty}\ \frac{1}{N}\log Z_{N}(\beta,h)=f(\beta,h)\ , \label{eq:GT2}
\end{equation}
does exist $J$-almost surely.
\end{theorem}

The proof is based on the fact that the fluctuations of the free energy per spin vanish exponentially fast as $N$ grows:
\begin{equation}
P\left(\Bigg |\frac{1}{\beta N}\log Z_{N}(\beta,h,J)-\frac{1}{\beta N}\ \mathbb{E}\log Z_{N}(\beta,h,J)\Bigg |\geq C\right)\leq\mathrm{e}^{-NC^2/2}\ .\label{eq:expfluct}
\end{equation}
Then, noticing that the r.h.s. of Eq. \eqref{eq:expfluct} is summable in $N$ for every fixed $C$, Borel-Cantelli lemma and the convergence given by \eqref{eq:GT1} imply \eqref{eq:GT2}.

\subsection{Exactness of the Broken Replica Symmetry solution}

After having established the existence of the thermodynamical limit in the SK model, we turn to the problem of proving the correctness of the Parisi mechanism for the phenomenon of spontaneous replica symmetry breaking.

The proof is based on two main ingredients. The first one is an interpolation method, invented and described in the marvellous paper by Guerra of 2003 \cite{Guerra}, which allows us to prove that the RSB ansatz is a rigorous lower bound for the quenched average of the free energy per spin, uniformly in the size of the system. The second idea is due to Talagrand, who made in 1998 the observation that, in order to prove an upper bound for the  quenched average free energy, it is sufficient to prove a lower bound on a similar quantity that involves 2 copies of the system (two real replicas). This observation was not very useful at that time, since there was not method to prove the lower bound. Soon after the discovery of the interpolating method, Talagrand combined Guerra's method of proving lower bounds, with his method to turn lower bounds into upper bounds, thus obtaining the proof.

We give here a brief sketch of the Guerra's interpolating method and we refer to the original work \cite{Guerra} for the details and to the work by Talagrand \cite{Talagrand} for a complete proof. 

Following Guerra we first formulate the RSB ansatz without using replicas. So let us consider the (convex) space $\aleph$ of the functional order parameter $x(q)$, as non-decreasing function of $q$, both $x$ and $q$ taking values on the interval $[0,1]$, i.e,
\begin{equation}
x:[0,1]\ni q\ \mapsto x(q)\in[0,1]\ , \ \ \  x\in\aleph\ .
\end{equation}  
It is useful to consider the case of piecewise constant functional order parameters, characterized by an integer $K$, and two sequences $q_0,q_1,\dots,q_K$ and $m_1,m_2,\dots,m_K$ of numbers satisfying
\begin{equation}
\begin{aligned}
&0=q_0\leq q_1\leq \dots\leq q_{K-1}\leq q_K=1,\\
&0\leq m_1\leq m_2\leq \dots\leq m_K\leq1,
\end{aligned}
\end{equation}
such that
\begin{equation}
x(q) = 
\begin{cases}
m_1\ \ \ \mathrm{for}\ \ \ q_0\leq q\leq q_1\ ,\\
m_2\ \ \ \mathrm{for}\ \ \ q_1\leq q\leq q_2\ ,\\
\dots\\
m_K\ \ \ \mathrm{for}\ \ \ q_{K-1}\leq q\leq q_K\ .
\end{cases}
\label{eq:piecewiseq}
\end{equation}
The replica symmetric case corresponds to 
\begin{equation}
K=2,\  q_1=\bar{q},\ m_1=0,\ m_2=1\ .
\end{equation}
The case $K=3$ is the first level of replica symmetry breaking, and so on. 

Let us now introduce the function $f_x(q,y)$ of the variables $q\in[0,1]$, $y\in\mathbb{R}$, depending also on the functional order parameter $x(q)$, defined as the solution of the non-linear antiparabolic equation
\begin{equation}
2\frac{\partial f}{\partial q}+\frac{\partial^2f}{{\partial y^2}}+x(q)\left(\frac{\partial f}{\partial y }\right)^2=0\, \label{eq:parabolic}
\end{equation} 
with final condition $f_x(1,y)=\log\cosh(\beta y)$. In the following we will omit, for simplicity, the dependence of $f$ from $\beta$.

It turns out that the function $f_x(q,y)$ is monotone in $x$, in the sense that $x(q)\leq \bar{x}(q)$ for all $0\leq q\leq 1$, implies $f_x(q,y)\leq f_{\bar{x}}(q,y)$ for any $0\leq q\leq 1$, $y\in \mathbb{R}$. Moreover $f_x(q,y)$ is pointwise continuous, so that, for generic $x,\bar{x}$, we have
\begin{equation}
|f_x(q,y)-f_{\bar{x}}(q,y)|\leq \frac{\beta^2}{2} \int_q^1|x(q')-\bar{x}(q')|\ {\mathrm d} q'\ .
\end{equation}
This result is very important. Indeed, any functional order parameter can be approximated through a piecewise constant one. The pointwise continuity allows us to deal mostly with piecewise constant order parameters.

Let us now define the trial \textit{free entropy} function, depending on the functional order parameter $x(q)$, as follows:
\begin{equation}
\bar{\alpha}_{x}(\beta,h)\equiv\log 2+f_x(0,h)-\frac{\beta^2}{2}\int_0^1{\mathrm d} q\ q\ x(q)\ .
\end{equation}
When multiplied by $-1/\beta$ this becomes the trial \textit{free energy} function. Notice that in this expression the function $f$ appears evaluated at $q=0$ and $y=h$, where $h$ is the value of the external magnetic field.

The Parisi RSB solution is defined by 
\begin{equation}
\bar{\alpha}_{\mathrm{RSB}}(\beta,h)\equiv\inf_{x}\ \bar{\alpha}_{x}(\beta,h)\ ,
\end{equation}
where the infimum is taken with respect to all functional order parameter $x(q)$. The main result of Guerra's work can be summarized in the following theorem

\begin{theorem}
\emph{(Guerra)}
\label{Guerra}
For all values of the inverse temperature $\beta$ and the external magnetic field $h$, and for any functional order parameter $x(q)$, the following bounds holds
\begin{equation}
\frac{1}{N}\ \mathbb{E}\log Z_{N}(\beta,h,J)\leq \bar{\alpha}_{x}(\beta,h) , \label{eq:GuerraLowerBound}
\end{equation}
uniformly in $N$. Consequently, we have also
\begin{equation}
\frac{1}{N}\ \mathbb{E}\log Z_{N}(\beta,h,J)\leq \bar{\alpha}_{\mathrm{RSB}}(\beta,h) , 
\end{equation}
uniformly in $N$. Moreover, for the thermodynamic limit, we have
\begin{equation}
\lim_{N\to\infty}\frac{1}{N}\ \mathbb{E}\log Z_{N}(\beta,h,J) \equiv\alpha(\beta,h)
\leq \bar{\alpha}_{\mathrm{RSB}}(\beta,h)\ ,
\end{equation}
and 
\begin{equation}
\lim_{N\to\infty}\frac{1}{N}\log Z_{N}(\beta,h,J) \equiv\alpha(\beta,h)
\leq \bar{\alpha}_{\mathrm{RSB}}(\beta,h)\ ,
\end{equation}
$J$-almost surely.
\end{theorem}
The proof of the theorem is long, and we refer to the original paper by Guerra for an exhaustive presentation. Here we only sketch the main ideas of the proof.

Consider a generic piecewise constant functional order parameter $x(q)$, as in \eqref{eq:piecewiseq}, and define the following interpolating partition function $\tilde{Z}(t;x)$
\begin{equation}
\begin{aligned}
&\tilde{Z}(t;x)\equiv\sum_{\{\sigma\}}\exp[\beta H_t(\sigma)]\\
&H_t(\sigma)=
\sqrt{\frac{t}{N}}\sum_{j>i=1}^NJ_{ij}\sigma_i\sigma_j+ h\sum_{i=1}^N\sigma_i+\sqrt{1-t}\sum_{a=1}^K\sqrt{q_a-q_{a-1}}\sum_{i=1}^N J_i^a\sigma_i
\ ,
\end{aligned}
\label{eq:InterpolZGuerra}
\end{equation}
where we have omitted the dependence of $\tilde{Z}(t;x)$ from $\beta$, $h$, $J$ and $N$. The numbers $J_i^a$ are additional independent centered unit Gaussian random variables, and the parameter $t$ runs in the interval $[0,1]$.

For $a=1,\dots,K$, let us call $\mathbb{E}_a$ the average with respect to all random variables $J_i^a$, $i=1,\dots,N$. Moreover, we call $\mathbb{E}_0$ the average with respect to all $J_{ij}$, and denote by $\mathbb{E}$ averages with respect to all $J$ random variables.

Now we define recursively the random variables $Z_0,Z_1,\dots,Z_K$
\begin{equation}
\begin{aligned}
&Z_K=\tilde{Z}(t;x)\ ,\\
&Z_{K-1}=\mathbb{E}_K Z_{K}\ ,\\
&\dots\\
&Z_0=\mathbb{E}_1 Z_{1}\ ,
\end{aligned}
\end{equation}
and the auxiliary function $\tilde{\alpha}_N(t)$
\begin{equation}
\tilde{\alpha}_N(t)=\frac{1}{N}\ \mathbb{E}_0\log Z_0\ .
\end{equation}
Due to the partial integrations, any $Z_a$ depends only on the $J_{ij}$ and on the $J_i^b$ with $b\leq a$, while in $\tilde{\alpha}_N(t)$ all $J$ have been completely averaged out.

At the extreme values of the interpolating parameter $t$ the function $\tilde{\alpha}_N(t)$ evaluates
\begin{equation}
\begin{aligned}
&\tilde{\alpha}_N(1)=\frac{1}{N}\ \mathbb{E}\log Z_N(\beta,h,J)\ ,\\
&\tilde{\alpha}_N(0)=\log 2+ f_q(0,h)\ .
\end{aligned}
\label{eq:boundaryAlpha}
\end{equation}

What we want to calculate is the $t$ derivative of $\tilde{\alpha}_N(t)$. Before to do this we need some few additional definitions. So let us introduce the random variables $f_a$, $a=1,\dots,K$:
\begin{equation}
f_a=\frac{Z_a^{m_a}}{\mathbb{E}_a(Z_a^{m_a})}\ ,
\end{equation}
and notice that they depend only on the $J_i^b$ with $b\leq a$, and are normalized: $\mathbb{E}(f_a)=1$. Following Guerra, we consider the $t$-dependent state $\langle\cdot\rangle$ associated with the Boltzmann factor in \eqref{eq:InterpolZGuerra}:
\begin{equation}
\langle\cdot\rangle=\frac{\sum_{\{\sigma\}}\ \cdot\ \exp[\beta H_t(\sigma)]}{\tilde{Z}(t;x)}
\end{equation}
and the replicated one
\begin{equation}
\langle\cdot\rangle^{(s)}=\frac{\sum_{\{\sigma^1\dots \sigma^s\}}\ \cdot\ \exp[\beta H_t(\sigma^1)+\dots \beta H_t(\sigma^s)]}{\left[\tilde{Z}(t;x)\right]^s}\ .
\end{equation}
Then we define the following states $\langle\cdot\rangle_a$, $a=0,\dots K$, as 
\begin{equation}
\begin{aligned}
&\langle\cdot\rangle_K=\langle\cdot\rangle\ ,\\
&\langle\cdot\rangle_a=\mathbb{E}_{a+1}\dots\mathbb{E}_{K}[f_{a+1}\dots f_K\ \langle\cdot\rangle]\ ,
\end{aligned}
\end{equation}
and the replicated ones as 
\begin{equation}
\begin{aligned}
&\langle\cdot\rangle_K^{(s)}=\langle\cdot\rangle^{(s)}\ ,\\
&\langle\cdot\rangle_a^{(s)}=\mathbb{E}_{a+1}\dots\mathbb{E}_{K}[f_{a+1}\dots f_K\ \langle\cdot\rangle^{(s)}]\ ,
\end{aligned}
\end{equation}
Finally, we define the $\overline{\cdot}^{\ a}$ averages as
\begin{equation}
\overline{\cdot}^{\ a} = \mathbb{E}\ \left[f_1\dots f_a\  \langle\cdot\rangle_a^{(s)}\right]\ .
\end{equation}
The idea behind the definition of the $\overline{\cdot}^{\ a}$ averages is the fact that they are able, in a sense, to concentrate the overlap fluctuations around the value $q_a$. 

With these definitions, one can show that the derivative of $\tilde{\alpha}_N(t)$ is given by
\begin{equation}
\frac{{\mathrm d} }{{\mathrm d} t}\tilde{\alpha}_N(t)=-\frac{\beta^2}{4}\left[1-\sum_{a=0}^{K}(m_{a+1}-m_a)\left(q_a^2-
\overline{(q_{12}-q_a)^2}^a\right)\ ,
\right]
\end{equation}
where $q_{12}$ is the overlap defined in \eqref{eq:overlap_ab}, i.e., $q_{12}=N^{-1}\sum_i\sigma_i^1\sigma_i^2$. 

By integrating with respect to $t$, and taking into account the boundary values \eqref{eq:boundaryAlpha}, we find the \textit{Guerra's sum rule}
\begin{equation}
\tilde{\alpha}_x(\beta,h)=\frac{1}{N}\mathbb{E}\log Z_{N}(\beta,h,J)+\frac{\beta^2}{4}\sum_{a=0}^{K}(m_{a+1}-m_{a})\int_{0}^{1}\overline{(q_{12}-q_a)^2}^a(t)\ {\mathrm d} t\ ,
\end{equation}
All terms in the sum are non-negative, since $m_{a+1}\geq m_{a}$, and the validity of the theorem is established.

\bibliography{references}
\bibliographystyle{unsrt}

\end{document}